\documentstyle[12pt]{article}
\begin{document}

\thispagestyle{empty}
\begin{flushright}
SLAC-PUB-7520\\
hep-ph/9705369\\
May 1997
\end{flushright}
\vspace*{1.5cm}
\centerline{\Large\bf CP Violation in Neutral Kaon Decays
\footnote{
Work supported by the Department of Energy under contract
DE-AC03-76SF00515.}}
\vspace*{1.5cm}
\centerline{{\sc G. Buchalla}}
\bigskip
\centerline{\sl Stanford Linear Accelerator Center}
\centerline{\sl Stanford University, Stanford, CA 94309, U.S.A.}

\vspace*{1.2cm}
\centerline{\bf Abstract}
\vspace*{0.2cm}
\noindent 
A brief review of the theoretical status of CP violation in
decays of neutral kaons is presented. We focus on three 
important topics: $\varepsilon$,
$\varepsilon'/\varepsilon$ and $K_L\to\pi^0\nu\bar\nu$.

\vspace*{2cm}
\centerline{\it Invited Talk presented at the Symposium on}
\centerline{\it Flavor-Changing Neutral Currents, Santa Monica,
California, February 19--21, 1997}

\vfill
\newpage
\pagenumbering{arabic}

\parindent=0pt

\section{Introduction}
%

The phenomenon of CP violation is of great current interest in
particle physics. It defines an absolute, physical distinction
between matter and antimatter and is a necessary condition for the
generation of a baryon asymmetry in the universe. Furthermore, 
studying CP violation tests our understanding of flavordynamics.
This sector of the Standard Model (SM) is clearly the most
complicated, involving spontaneous breaking of electroweak
symmetry and containing most of the free parameters of the model,
including quark masses, CKM angles and the CP violating phase.
In order to address these fundamental questions experimentally,
neutral kaons have proved to be a key tool. In fact, until today
CP violation has been exclusively observed in a few decay modes of the
long lived neutral kaon ($K_L\to\pi\pi$, $\pi l\nu$, 
$\pi^+\pi^-\gamma$). All of the observed effects are accounted for
by a single complex parameter $\varepsilon$, consistent with CP
violation in the mass matrix only. The unambiguous demonstration
of direct CP violation in $K_L\to\pi\pi$, measured by
$\varepsilon'/\varepsilon$, is so far still elusive. The topic is
currently still under active investigation. Beyond that, the field of
neutral kaon CP violation will continue to offer interesting 
opportunities in the future. Among the possibilities that have been
discussed are the rare decay $K_L\to\pi^0e^+e^-$, muon polarization
in $K_L\to\mu^+\mu^-$, and in particular the `gold-plated' mode
$K_L\to\pi^0\nu\bar\nu$. The latter is theoretically extremely clean
and offers excellent prospects for high precision flavor physics.
In the following we will briefly summarize the theoretical status
of $\varepsilon$, $\varepsilon'/\varepsilon$ and 
$K_L\to\pi^0\nu\bar\nu$, three main topics
in the study of CP violation with neutral kaons.
A more complete account and detailed references may be found
in \cite{BBL}.

\section{Indirect CP Violation in $K^0\to\pi\pi$: $\varepsilon$} 

The parameter $\varepsilon$ is determined by the imaginary part
of the element $M_{12}$ in the neutral kaon mass matrix, 
which in turn is generated by the usual
$\Delta S=2$ box-diagrams. The low energy effective Hamiltonian
contains a single operator $(\bar ds)_{V-A}(\bar ds)_{V-A}$
in this case and one obtains
\begin{equation}\label{epsth}
\varepsilon=e^{i\frac{\pi}{4}}
\frac{G^2_F M^2_W f^2_K m_K}{12\pi^2\sqrt{2}\Delta M_K}B_K\cdot\mbox{Im}
\left[\lambda^{*2}_c S_0(x_c)\eta_1+\lambda^{*2}_t S_0(x_t)\eta_2
  +2\lambda^*_c\lambda^*_t S_0(x_c,x_t)\eta_3\right]
\end{equation}
Here $\lambda_i=V^*_{is}V_{id}$, $f_K=160\,MeV$ is the kaon decay
constant and the bag parameter $B_K$ is defined by
\begin{equation}\label{bkrsi}
B_K=B_K(\mu)[\alpha^{(3)}_s(\mu)]^{-2/9}
 \left[1+J_3 \alpha^{(3)}_s(\mu)/(4\pi)\right]
\end{equation}
where
$\langle K^0|(\bar ds)_{V-A}(\bar ds)_{V-A}|\bar K^0\rangle
  \equiv 8/3 B_K(\mu) f^2_K m^2_K$.
The index $(3)$ in eq. (\ref{bkrsi}) refers to the number of flavors
in the effective theory and $J_3=307/162$ (in the NDR scheme).
\\
The Wilson coefficient multiplying $B_K$ in (\ref{epsth}) consists
of a charm contribution, a top contribution and a mixed top-charm
contribution. It depends on the quark masses, $x_i\equiv m^2_i/M^2_W$,
through the functions $S_0$. The $\eta_i$ are the corresponding
short-distance QCD correction factors (which depend only slightly on
quark masses).
Numerical values for $\eta_1$, $\eta_2$ and $\eta_3$ are summarized
in Table \ref{etaitab}.
\begin{table}
\centering
\caption{ \it NLO results for $\eta_i$ with
$\Lambda^{(4)}_{\overline{MS}}=(325\pm 110)\,MeV$,
$m_c(m_c)=(1.3\pm 0.05)\,GeV$, $m_t(m_t)=(170\pm 15)\,GeV$.
The third column shows the uncertainty due to the errors in
$\Lambda_{\overline{MS}}$ and quark masses. The fourth column
indicates the residual renormalization scale uncertainty at NLO
in the product of $\eta_i$ with the corresponding mass dependent function
from eq. (1). The central values of the
QCD factors at LO are also given for comparison.
}
\vskip 0.1 in
\begin{tabular}{|c|c|c|c|c|c|} \hline
& NLO(central) & $\Lambda_{\overline{MS}}$, $m_q$ &
scale dep. & NLO ref. & LO(central) \\
\hline
\hline
$\eta_1$ & 1.38 & $\pm 35\%$ & $\pm 15\%$ & \cite{HN1} & 1.12 \\
\hline
$\eta_2$ & 0.574 & $\pm 0.6\%$ & $\pm 0.4\%$ & \cite{BJW} & 0.61 \\
\hline
$\eta_3$ & 0.47 & $\pm 3\%$ & $\pm 7\%$ & \cite{HN3} & 0.35 \\
\hline
\end{tabular}
\label{etaitab}
\end{table}

$\varepsilon$ is dominated by the top contribution ($\sim 70\%$).
It is therefore rather satisfying that the related short distance
part $\eta_2 S_0(x_t)$ is theoretically extremely well under
control, as can be seen in Table \ref{etaitab}. Note in
particular the very small scale ambiguity at NLO, $\pm 0.4\%$
(for $100\,GeV\leq\mu_t\leq 300\,GeV$). This intrinsic theoretical
uncertainty is much reduced compared to the leading order result
where it would be as large as $\pm 9\%$.
\\
The $\eta_i$ factors and the hadronic matrix element are not
physical quantities by themselves. When quoting numbers it is therefore
essential that mutually consistent definitions are employed.
The factors $\eta_i$ described here are to be used in conjunction
with the so-called scheme- (and scale-) invariant bag parameter
$B_K$ introduced in (\ref{bkrsi}). The last factor on the rhs of
(\ref{bkrsi}) enters only at NLO. As a numerical example, if the
(scale and scheme dependent) parameter $B_K(\mu)$ is given in the
NDR scheme at $\mu=2GeV$, then (\ref{bkrsi}) becomes
$B_K=B_K(NDR,2\,GeV)\cdot 1.31\cdot 1.05$.
\\
The quantity $B_K$ has to be calculated by non-perturbative
methods. Large $N_C$ expansion techniques for instance find
values $B_K=0.75\pm 0.15$. 
The results obtained in other approaches are reviewed in \cite{BBL}.
Ultimately a first
principles calculation should be possible within lattice gauge
theory. Ref. \cite{SHA} quotes an estimate of
$B_K(NDR,2\,GeV)=0.66\pm 0.02\pm 0.11$ in full QCD. The first error
is the uncertainty of the quenched calculation. It is quite small
already and illustrates the progress achieved in controlling
systematic uncertainties in lattice QCD \cite{SHA,JLQCD}. 
The second error represents the
uncertainties in estimating the effects of quenching and 
non-degenerate quark masses.
\\
Phenomenologically $\varepsilon$ is used to determine the CKM
phase $\delta$.
The relevant input parameters are
$B_K$, $m_t$, $V_{cb}$ and $|V_{ub}/V_{cb}|$. 
A typical analysis of constraints on CKM parameters from $\varepsilon$
can be found for instance in \cite{BBL}.

\section{Direct CP Violation in $K^0\to\pi\pi$: $\varepsilon'/\varepsilon$}

The theoretical expression for $\varepsilon'/\varepsilon$
can be written as
\begin{equation}\label{epeapr}
\frac{\varepsilon'}{\varepsilon}=
\frac{\omega G_F}{2|\varepsilon|\mbox{Re} A_0}\mbox{Im}\lambda_t
\left(y_6\langle Q_6\rangle_0-\frac{1}{\omega}y_8\langle Q_8\rangle_2
+\ldots\right)
\end{equation}
where, for the purpose of illustration we kept only the numerically
dominant terms.
Here $y_i$ are Wilson coefficients,
$\langle Q_i\rangle_{0,2}\equiv\langle\pi\pi(I=0,2)|Q_i|K^0\rangle$,
$A_{0,2}$ are the $K^0\to\pi\pi(I=0,2)$ amplitudes and 
$\omega=\mbox{Re}A_2/\mbox{Re}A_0$.
The operator
$Q_6$ originates from gluonic penguin diagrams and $Q_8$ from
electroweak contributions. The matrix elements of $Q_6$ and
$Q_8$ have the form
$\langle Q_6\rangle_0\sim B_6/m^2_s$ and
$\langle Q_8\rangle_2\sim B_8/m^2_s$, where $B_6$ and $B_8$ are 
bag parameters.
$y_6\langle Q_6\rangle_0$ and $y_8\langle Q_8\rangle_2$
are positive. The value for $\varepsilon'/\varepsilon$
in (\ref{epeapr}) is thus characterized by a cancellation of
competing contributions. Since the second contribution is an
electroweak effect, suppressed by $\sim\alpha/\alpha_s$ compared
to the leading gluonic penguin $\sim\langle Q_6\rangle_0$,
it could appear at first sight that it should be altogether
negligible for $\varepsilon'/\varepsilon$. However, a number of
circumstances actually conspire to systematically enhance the
electroweak effect so as to render it a very important contribution.
First,
unlike $Q_6$, which is a pure $\Delta I=1/2$ operator,
$Q_8$ can give rise to the $\pi\pi(I=2)$ final state and thus
yield a nonvanishing isospin-2 component in the first place.
Second,
the ${\cal O}(\alpha/\alpha_s)$ suppression is largely compensated
by the factor $1/\omega\approx 22$ in (\ref{epeapr}), reflecting the
$\Delta I=1/2$ rule.
Third,
$\langle Q_8\rangle_2$ is somewhat enhanced relative to
$\langle Q_6\rangle_0$, which vanishes in
the chiral limit.
Finally,
$-y_8\langle Q_8\rangle_2$ gives a negative contribution to
$\varepsilon'/\varepsilon$ that strongly grows with $m_t$
\cite{FR,BBH}. For the actual top mass value it is
quite substantial.

The Wilson coefficients $y_i$ have been calculated at NLO
\cite{BJLW,CFMR}. The short-distance part is therefore quite
well under control. The remaining problem is then the computation
of matrix elements, in particular 
$\langle Q_6\rangle_0$ and $\langle Q_8\rangle_2$. The cancellation
between their contributions enhances the relative sensitivity of
$\varepsilon'/\varepsilon$ to the anyhow uncertain hadronic parameters
which makes a precise calculation of $\varepsilon'/\varepsilon$
impossible at present. 
In a recent analysis Buras et al. \cite{BJL96} find for
$(\varepsilon'/\varepsilon)/10^{-4}$
\begin{equation}\label{epe1}
7.4\pm 8.6\ (s)\qquad 3.6\pm 3.4\ (g)\qquad m_s(2 GeV)=(129\pm 18)MeV
\end{equation}
\begin{equation}\label{epe2}
21.5\pm 21.5\ (s)\qquad 10.4\pm 8.3\ (g)\qquad m_s(2 GeV)=(86\pm 17)MeV
\end{equation}
Here $(g)$ refers to the assumption of a
Gaussian distribution of errors in the input parameters,
$(s)$ to the more conservative `scanning' of parameters over their
full allowed ranges.
The lower values for the strange quark mass $m_s$ in (\ref{epe2})
correspond to recent lattice results \cite{GBH,GOU}.
Within the rather large uncertainties (\ref{epe2}) is compatible with 
experiment, which gives $(23\pm 7)\cdot 10^{-4}$ (CERN-NA31) and
$(7.4\pm 5.9)\cdot 10^{-4}$ (FNAL-E731).
On the other hand, (\ref{epe1}) is consistent with E731,
but somewhat low compared to NA31.
Similar results have been obtained by other authors
\cite{CFMRS,BEF2,HEI}.

In conclusion, the SM prediction for $\varepsilon'/\varepsilon$
suffers from large hadronic uncertainties, reinforced by substantial
cancellations between the $I=0$ and $I=2$ contributions. Despite this
problem, the characteristic pattern of CP violation observed in
$K\to\pi\pi$ decays, namely $\varepsilon={\cal O}(10^{-3})$ and
$\varepsilon'={\cal O}(10^{-6})$ (or below), is well accounted for
by the standard theory, which can be considered a non-trivial success
of the model.
\\
On the experimental side a clarification of the current situation is to 
be expected from the upcoming new round of $\varepsilon'/\varepsilon$
experiments conducted at Fermilab (E832), CERN (NA48) and
Frascati (KLOE). The goal is a measurement of $\varepsilon'/\varepsilon$
at the $10^{-4}$ level. The demonstration that $\varepsilon'\not= 0$
would constitute a qualitatively new feature of CP violation and
as such be of great importance. 
However, due to the large uncertainties in the
theoretical calculation, a quantitative use of this result for the
extraction of CKM parameters will unfortunately be severely limited.
For this purpose one has to turn to theoretically cleaner
observables.

\section{$K_L\to\pi^0\nu\bar\nu$}
%

The rare decay $K_L\to\pi^0\nu\bar\nu$ is a very attractive
probe of flavordynamics.
In particular, $K_L\to\pi^0\nu\bar\nu$ is a
manifestation of {\em large direct CP violation\/} in the SM.
A small effect from indirect CP violation
related to the kaon $\varepsilon$-parameter contributes
below $\sim 1\%$ in the branching ratio and is therefore negligible.
\\
In addition to having this phenomenologically interesting feature,
$K_L\to\pi^0\nu\bar\nu$ can be calculated as a function of fundamental
SM parameters with exceptionally small  theoretical error.
The main reasons are the hard GIM suppression of long distance
contributions \cite{RS}, and the semileptonic character, which allows
us to extract the hadronic matrix element
$\langle\pi^0|(\bar sd)_V|K^0\rangle$ from $K^+\to\pi^0e\nu$ decay
using isospin symmetry. As a consequence $K_L\to\pi^0\nu\bar\nu$ is
based on a purely short-distance dominated flavor-changing neutral
current, which is reliably calculable in perturbation theory. The
CP properties help to further improve the theoretical accuracy, rendering 
even the charm contribution completely negligible so that the clean
top contribution fully dominates the decay.
Next-to-leading QCD effects have been calculated and reduce
the leading order scale ambiguity of $\sim \pm 10\%$ to an
essentially negligible $\sim \pm1\%$ \cite{BB2}.
Isospin breaking corrections in the extraction of the matrix element
have also been evaluated. They lead to an overall reduction of the
branching ratio by $5.6\%$ \cite{MP}.
As a result of all these developments, the theoretical uncertainty
in $K_L\to\pi^0\nu\bar\nu$ is safely below $2\%$.
\\
The quantity $B(K_L\to\pi^0\nu\bar\nu)$ offers probably the
best accuracy in determining $\mbox{Im} V^*_{ts}V_{td}$ or,
equivalently, the Jarlskog parameter
$J_{CP}=\mbox{Im}(V^*_{ts}V_{td}V_{us}V^*_{ud})$.
The prospects here are even better than for $B$ physics at the LHC,
assuming a $\pm 10\%$ measurement of $B(K_L\to\pi^0\nu\bar\nu)$
at about the central value of SM predictions \cite{BB96}.
The SM expectation for the branching ratio \cite{BJLnew} is 
$(2.8\pm 1.7)\cdot 10^{-11}$, where the uncertainty is due to our
imprecise knowledge of CKM parameters. 
The current upper bound
from direct searches \cite{WEA} is $5.8\cdot 10^{-5}$. An indirect
upper bound, using the current limit on $B(K^+\to\pi^+\nu\bar\nu)$
\cite{ADL} and isospin symmetry, can be placed \cite{GN} at 
$1.1\cdot 10^{-8}$.
Several activities are under way aiming for an actual measurement of
$K_L\to\pi^0\nu\bar\nu$. A proposal exists at
Brookhaven (BNL E926) to measure this decay at the AGS 
with a sensitivity of ${\cal O}(10^{-12})$.
There are furthermore plans to pursue this mode with comparable
sensitivity at Fermilab and KEK. More details can be found in the
contributions by D. Bryman, T. Nakaya, K. Arisaka and T. Inagaki
to these proceedings.

\section{Summary}
%

Decays of neutral kaons provide the only instance where CP violation
has been observed to date. The quantity $\varepsilon$
measures indirect CP violation, it is experimentally very precisely
known and leads to important constraints on CKM parameters.
The search for direct CP violation in $K_L\to\pi\pi$, measured by
$\varepsilon'/\varepsilon$, is still ongoing. The SM prediction for
this quantity is plagued by large hadronic uncertainties, which 
severely limits the possibility of extracting useful information on
CKM quantities from this observable. 
Eventually, a measurement of $K_L\to\pi^0\nu\bar\nu$ would
open up exciting prospects for precision studies in flavor physics,
complementary and competitive to CP violation studies in B decays.
Theoretical progress has been achieved on all three topics,
$\varepsilon$, $\varepsilon'/\varepsilon$ and $K_L\to\pi^0\nu\bar\nu$,
in particular through the calculation of next-to-leading order
QCD effects.

The study of CP violation in neutral kaon decays has yielded crucial
insight into fundamental physics in the past and it is still under
active investigation at present. 
Excellent opportunities, as those provided by $K_L\to\pi^0\nu\bar\nu$,
continue to exist for the future.

\end{document}